\documentclass[sigconf]{acmart}

\usepackage[utf8]{inputenc}

\usepackage{latexsym}
\usepackage{pdflscape}
\usepackage{afterpage}
\usepackage{epsfig}
\usepackage{adjustbox}
\usepackage{amsmath}

\hypersetup{
  bookmarks=false
}

\usepackage[ruled,vlined,linesnumbered]{algorithm2e} 
\usepackage{balance}

\usepackage{booktabs} 
\usepackage{xspace}
\usepackage{url}
\usepackage{wasysym}
\usepackage{enumerate}
\usepackage{multirow}
\usepackage{array}
\usepackage{lscape}

\newcommand{\tick}{$\newmoon$}
\newcommand{\fail}{\Circle}

\newtheorem{propos}{Property}[section]
\newtheorem{coroll}[propos]{Corollary}

\newtheorem{assum}{Assumption}

\usepackage{soul}
\setstcolor{red}
\setul{}{.2ex}

\newcommand{\D}{\mathcal{D}}

\newcommand{\eff}[0]{\text{OIE}}

\newcommand{\obsd}[2]{\mathcal{O}_{#1}(#2)}

\newcommand{\IQf}[1]{{\mathcal{I}}_{#1}}
\newcommand{\IQd}[2]{\mathcal{I}_{#1}\big(#2\big)}

\newcommand{\E}[1]{H\big(#1\big)}

\copyrightyear{2018}
\acmYear{2018}
\setcopyright{acmlicensed}
\acmConference[ICTIR '18]{2018 ACM SIGIR International Conference
on the Theory of Information Retrieval}{September 14--17,
2018}{Tianjin, China}
\acmBooktitle{2018 ACM SIGIR International Conference on the Theory
of Information Retrieval (ICTIR '18), September 14--17, 2018, Tianjin,
China}
\acmPrice{15.00}
\acmDOI{10.1145/3234944.3234958}
\acmISBN{978-1-4503-5656-5/18/09}

\title[A Formal Account of  Effectiveness Evaluation and Ranking Fusion]{A Formal Account of \\ Effectiveness Evaluation and Ranking Fusion}

\author{Enrique Amig{\'o}}
\affiliation{%
  \institution{UNED NLP \& IR Group}
  \city{Madrid}
  \country{Spain}} 
\email{enrique@lsi.uned.es}

\author{Fernando Giner}
\affiliation{%
  \institution{UNED NLP \& IR Group}
  \city{Madrid}
  \country{Spain}} 
\email{fginer3@alumno.uned.es}

\author{Stefano Mizzaro}
\affiliation{%
  \institution{University of Udine}
  \city{Udine}
  \country{Italy}} 
\email{mizzaro@uniud.it}

\author{Damiano Spina}
\affiliation{%
  \institution{RMIT University}
  \city{Melbourne}
  \country{Australia}} 
\email{damiano.spina@rmit.edu.au}

\begin{document}

\begin{abstract}
This paper proposes a theoretical  framework which models the  information provided by retrieval systems in terms of Information Theory. 
The proposed framework allows to formalize: (i) system effectiveness as an information theoretic similarity between system outputs and human assessments, and (ii) ranking fusion as an information quantity measure. As a result, the proposed effectiveness metric improves popular metrics in terms of formal constraints. In addition, our empirical experiments suggest that it captures quality aspects from traditional metrics, while the reverse is not true. 
Our work also advances the understanding of theoretical foundations of the  empirically known phenomenon of effectiveness increase when  combining retrieval system outputs in an unsupervised manner.
\end{abstract}

%
%

\ccsdesc[500]{Information systems~Information retrieval}

\keywords{information theory; evaluation; ranking fusion}

\maketitle

\section{Introduction}
\label{sec:introduction}

Most of the research in the field of Information Retrieval (IR) is empirically-based. The effectiveness of retrieval approaches are typically validated over large data sets, most of them developed in the last decade. The effectiveness of ranking fusion and learning-to-rank algorithms are also validated in an empirical way. 
In addition, effectiveness metrics are supported by empirical user behavior studies or meta-metrics such as robustness or sensitivity~\cite{Golbus-13,Amigo_2013}.

On the other hand, some works aim to provide explanations for some phenomena observed in empirical experiments. For instance, the Probability Ranking Principle~\cite{rijbsbergen1979ir} assumes that retrieval systems return documents ranked in order of decreasing probability of relevance to the user. In the same way, based on empirical user behavior observations, evaluation metrics are supported by the top-heaviness principle~\cite{busin13}, which gives more weight to highly-ranked documents in the evaluation process. Studies in unsupervised ranking fusion algorithms have reported empirically that the most effective combinations of rankings are those in which the relevant documents are unanimously early-ranked, while the retrieved non-relevant documents vary across rankings~\cite{Lee-95,Vogt-98}. Likewise, other studies have reported empirically that human assessments can be replaced successfully --at least to some extent-- by the average of system outputs in an evaluation campaign~\cite{Soboroff-01,Aslam-03}. 

In this paper, we aim to define
a theoretical framework which models the phenomena described above. The framework is based on the notion of \emph{Observational Information Quantity}: Rather than focusing on document content, this framework  models  the  information provided by retrieval systems (document rankings) and human assessors in terms of Information Theory~\cite{shannon1949mathematical}.
On the basis of observational information quantity we then define an entropy-like notion that allows the formalization of system effectiveness as an 
information-theoretic similarity between system outputs and human assessments. The proposed framework also models ranking fusion as an information quantity measure.  

The resulting effectiveness metric improves most of existing metrics in terms of formal constraints. In other words, the proposed framework gives a basis --grounded in Information Theory-- for  effectiveness metrics, which were traditionally supported by user behavior modeling.  Additionally, our  experiments corroborate this analysis, showing  that the proposed metric captures quality aspects from traditional metrics, while the reverse is not true.

On the other hand, our work provides a theoretical foundations of the  empirically-known phenomenon of 
effectiveness increase when  combining retrieval system outputs in an unsupervised manner. Our experiments also check empirically the assumptions in which the proposed theoretical framework is grounded.

Let us remark that this work does not attempt to provide \textit{better} solutions than those presented in previous work; rather we aim at defining a global theoretical framework on which to base future improvements.

The rest of the paper is organized as follows. Section~\ref{sec:relatedwork} discusses related work and Section~\ref{sec:theory} introduces the theoretical framework based on Observational Information Quantity. Section~\ref{sec:evaluation} analyzes how our proposed framework can be used to inform an effectiveness evaluation metric that satisfies a set of formal constraints.
Section~\ref{sec:fusion} describes the justification of ranking fusion based on Observational Information Quantity. 
Section~\ref{sec:derivation} connects the definitions of our framework with those of the classical Information Theory by Shannon.
Finally, Section~\ref{sec:conclusions} concludes the work.

\section{Related Work}
\label{sec:relatedwork}


\subsection{Measuring Effectiveness}

Most of current metrics estimate effectiveness by assuming an underlying user model for browsing relevant and non-relevant documents returned in the system output ranking. For instance,  Discount Cumulative Gain~\cite{DCG} assumes that the probability of exploring deeper ranking positions decreases in a logarithmic manner.  Expected Reciprocal Rank (ERR)~\cite{Chapelle-09} assumes a cascade model in which the user is looking for a particular document.  Rank-Biased Precision (RBP)~\cite{RBP} assumes that a fixed probability of exploring the  next document in the ranking. According to the analysis by \citet{Amigo_2013} none of the most popular metrics satisfies completely a set of five formal constraints. RBP satisfies four of them but not the \textit{confidence} constraint, which penalizes the addition of non-relevant documents at the end of the ranking.

Some authors have focused on  explaining  evaluation metrics in terms of Measurement Theory, tackling the issue of the suitability of the interval scale assumption~\cite{ferrante2018general} or interpreting metrics as an homomorphism (measurement) between effectiveness and systems~ \cite{Ferrante:2015}. These works state formal constraints and desirable properties, but they do not derive any particular approach.

In this paper, we  apply an information theory-based similarity measure to compare system outputs against the gold-standard. Our theoretical analysis shows that user behavior based constraints can be satisfied by grounding the metric in information theory principles.

\subsection{Ranking Fusion}
 Finding a theoretical explanation for the effectiveness of combining system outputs in an unsupervised manner has been largely explored in the literature. This problem has been modeled from two closely-related perspectives: classifier ensembles and  ranking fusion. 
 
From the first perspective,  the literature shows that combining classifiers is effective when the individual classifiers are accurate and diverse. \citet{Hansen-90} proved that if the average error rate for an example is less than 50\% and the component classifiers in the ensemble are independent in the production of their errors, the expected error for that example can be reduced to zero as the number of classifiers combined goes to infinity. This theoretical analysis was actually reported by the Condorcet's jury theorem in 1785~\cite{condorcet1785, austen-smith_banks_1996}. However, such assumptions rarely hold in practice. \citet{Krogh-95} later  formally showed  that an ideal ensemble consists of highly correct classifiers that disagree as much as possible. In general, a point of consensus is that when the classifiers make statistically independent errors, the combination has the potential to increase the performance of the system. Other studies assume correlation between signals, but equal performance and homogeneous correlation~\cite{Kaniovski-11}, which is also non-realistic in the context of information systems. \citet{Matan-96} analyzed the upper and lower bounds of classification of a majority based ensemble. In the particular context of information retrieval tasks,  \citet{Fox-94}  found that the best  combination  strategy  consisted  of summing  the  outputs  of the  retrieval algorithms, and \citet{Hull-96} found that the best improvement in performance in the context of a filtering task came from the simple averaging strategy.

From the ranking fusion perspective, \citet{MontagueA02} reported an important improvement of unsupervised combined systems w.r.t. the best single system in multiple TREC test beds.  In addition, just like in the classification scenario, the need for avoiding redundant systems has been reported in the context of ranking fusion. For instance, \citet{Nuray-06} reported effectiveness improvement when selecting rankings that differ from the majority voting in the ranking fusion process. \citet{Lee-95} and later \citet{Vogt-98} found that the best combinations were between systems that retrieve similar sets of relevant documents and dissimilar sets of non-relevant documents.  There exist other works that reformulate ranking fusion algorithms in terms of probability estimations, always under the independence assumption~\cite{bah2016probabilistic,Anava-16,markovits2012predicting}. Finally, \citet{Amigo-17} proposed an extension of the notion of Information Quantity in order to generalize different ranking fusion methods in the context of text similarity. In this paper, we review this notion extending it to \emph{observational entropy}.

\section{Observational Information Quantity}
\label{sec:theory}

\subsection{An Example}

Let us start with a simple example that considers the output of a set of information retrieval systems as in Table~\ref{tab:example}. 
The collection $\D$ contains a large amount of documents. 
There are three systems $s_1$, $s_2$, and $s_3$
that return different documents producing rankings $r_1$, $r_2$, and $r_3$ of length $3$. 
We assume that documents out of these rankings share the same infinite rank. 

\begin{table}[tbp]
\centering
\caption{Example of system outputs and human assessments.\label{tab:example}}
\small
\begin{tabular}{ccccc}
\toprule
 \multirow{2}{*}{Rank}    & \multicolumn{4}{c}{$\Gamma$} \\
\cmidrule{2-5}
& $r_1$ & $r_2$ & $r_3$ & Human assessments $g$ \\
\midrule
1 & $r_1(d_1)$ &  $r_2(d_3)$ & $r_3(d_3)$ &  \multirow{2}{*}{$g(d_1)=g(d_4)=1$}  \\
2 & $r_1(d_2)$ &  $r_2(d_1)$ & $r_3(d_1)$ &  \multirow{4}{*}{$g(d_{i=2,3,5,6,\ldots})=0$}  \\
3 & $r_1(d_4)$ & $r_2(d_2)$ &$r_3(d_2)$ &   \\
\cmidrule{1-4}
n/a ($\infty$) & \multicolumn{3}{c}{$r_{1,2,3}(d_5, d_6,\ldots) = 0,\hspace{1ex}r_{2,3}(d_4)=0$ } &  \\
\bottomrule
\end{tabular}
\end{table}

Rather than considering the content of the documents, we \textit{observe} the documents from the perspective of a set of information retrieval systems. The first step should be being able to \textit{measure} the quantity of information provided by systems or human assessments for single documents. This information measurement should satisfy the following three properties. 

First, the more a document is highly ranked according to a retrieval system, or relevance scored according to human assessors, the more the document is discriminated against the large collection (increasing  informativeness). For instance, according to $r_1$ in the example, we have more information about the relevance of $d_1$ than $d_2$.  

Second, the earlier documents are ranked according to different systems, the more information about their relevance we observe. This is in line with the conclusions found in previous work~\cite{Lee-95,Vogt-98}: the most effective combinations of rankings are those in which the relevant documents are unanimously early ranked, while the retrieved non-relevant documents vary across rankings. For instance, $d_1$ occurs at the first or second position in every ranking. Therefore, we have more information to estimate the relevance of $d_1$ than other documents. 

Third, redundant systems provide less information than non-redundant systems. In relation to this, the profits of combining non-redundant systems have been reported in both the ranking~\cite{Nuray-06} and classification~\cite{Krogh-95} scenarios. For instance, $d_3$ is ranked in the first position by $r_2$ and $r_3$, but both rankings seems to be similar, that is, they seem to be providing the same information.

\subsection{The Framework}

Let us consider a set of \textit{relevance signals}, $\Gamma$, which consists of the set of rankings and human assessments, $\Gamma=\{r_1,\ldots,r_n,g\}$\footnote{We will use $\gamma \in \Gamma$ to refer to signals in a general manner.}, which assign scores to documents in a collection of documents, $\D$. Then, we  define \emph{unanimous outscoring} as follows.\footnote{We assume that earlier rank positions correspond with higher scores.}

\begin{definition}
\label{def:outscoring}
\textit{A document, $d$, is  unanimously outscored by another document, $d'$,  according to a set of signals, $\Gamma$, whenever it is outscored for every signal simultaneously:} 

\begin{equation*}
d' \ge_\Gamma d \Longleftrightarrow \forall \gamma \in \Gamma \ldotp \gamma(d') \ge \gamma(d) .
\end{equation*}
\end{definition}
In the following we will use both $\le_\Gamma$ and $\ge_\Gamma$, with the obvious meaning. Going back to the example in Table~\ref{tab:example}, we obtain the following outscoring relationships regarding to $d_1, \ldots, d_4$. Being $\Gamma=\{r_1,r_2,r_3, g\}$ (note that we also take into account the human assessments in the ground truth $g$):
\begin{equation}\label{eq:outscoringex}
\begin{aligned}
   d_1  \ge_\Gamma & d_1 & 
 d_1, d_2 \ge_\Gamma  &  d_2 \\
  d_3 \ge_\Gamma & d_3 & 
 d_1, d_4 \ge_\Gamma &   d_4 .
\end{aligned}
\end{equation}

Documents $d_1$ and $d_3$ are only outscored by themselves, i.e., there is no other documents that is unanimously ranked earlier in $\Gamma$. 
$d_2$ is outscored by $d_2$ and $d_1$  (including $g$). Likewise, $d_4$ is outscored by itself and also by $d_1$, given that it is corroborated by the three rankings and the gold $g$. 

Then, the observational information quantity of a document is defined as follows.

\begin{definition}
  \label{def:FPIQ}
\textit{
The Observational Information Quantity, $\IQd{\Gamma}{d}$, of a document, $d$, under a set of signals, $\Gamma$, is the minus logarithm of the probability of being unanimously outscored by other documents:
\begin{equation}\label{eq:OIQ}
\IQd{\Gamma}{d}= - log\left(P_{d'\in \D}\big(d' \ge_\Gamma d\big)\right).
\end{equation}
}
\end{definition}

In other words, the more a document is unanimously outscored simultaneously in all signals by other documents, the less the document is informative. In consequence,  highly informative documents are those that are highly scored by all rankings and the human assessment.

For instance, going back to the example in Table~\ref{tab:example} and taking into account inequalities in Eq.~\eqref{eq:outscoringex}, the Observational Information Quantity of document documents is:\footnote{Note that the probabilities are computed as frequencies.}
\begin{align*}
&\IQd{\Gamma}{d_1}=-\log\left(\frac{1}{|\D|}\right), 
&\IQd{\Gamma}{d_2}=-\log\left(\frac{2}{|\D|}\right), \\
&\IQd{\Gamma}{d_3}=-\log\left(\frac{1}{|\D|}\right), 
&\IQd{\Gamma}{d_4}=-\log\left(\frac{2}{|\D|}\right).
\end{align*}

Documents that obtain the lowest score by all signals (i.e., documents that are both non-retrieved and non-relevant) obtain the lowest Observational Information Quantity, as they are outscored by $d_1$, $d_2$, $d_3$, $d_4$ and by themselves:
\begin{equation*}
\forall i \in \{5, 6, \ldots \} \ldotp  \IQd{\Gamma}{d_{i}} = -\log\left(\frac{|\D|}{|\D|}\right)=0.
\end{equation*}

Our formalization of observational information quantity matches with the definition provided by \citet{Amigo-17} for similarity measures fusion. Here, we  extend it to define an entropy-like notion.

\begin{definition}
  \label{def:H}
\textit{
  The Observational Entropy, $\E{\Gamma}$, of a set  of signals, $\Gamma$, is the expected observational information quantity across the document set, $\D$:
\begin{equation*}
\E{\Gamma}=\frac{\sum_{d\in\D}\IQd{\Gamma}{d}}{|\D|}.
\end{equation*}
}
\end{definition}

Intuitively, the observational entropy of a set of signals represents the extent to which finding unanimous improvement is unlikely. Thus, non-correlated signals will tend to achieve a higher entropy. Also note that this definition is inspired by the classical Shannon's Entropy, but it differs from the original because it uses two different probability distributions: the probability distribution of outscoring (the one used in Equation~\eqref{eq:OIQ}) and the probability distributions of a document ($\frac{1}{|\mathcal{D}|}$).  Section~\ref{sec:derivation} explains the connection between traditional and observational information quantity. For the sake of simplicity, we will denote hereafter the entropy for a single signal set as $\E{\gamma}$.

\subsection{Properties}

Observational Information Quantity and Observational Entropy satisfy the following general properties, that will be useful in the following.
\footnote{See formal proofs in Appendix~\ref{sec:appendix}.}

\begin{propos}
\label{prop:ValMon} For all $\gamma \in \Gamma$, $\IQd{\Gamma}{d}$ 
is monotonic w.r.t. signal values $\gamma(d)$: 
\begin{equation*}
d_1 \ge_\Gamma d_2 \Longrightarrow
\IQd{\Gamma}{d_1} \ge
\IQd{\Gamma}{d_2} .
\end{equation*}

\end{propos}

Property~\ref{prop:ValMon} implies the following corollary.
\begin{coroll}
\label{prop:SingleSignal}
 The observational information quantity of a document  under a single signal grows with its signal value:
\begin{equation*}
\IQd{\{\gamma\}}{d} \propto \gamma(d) .
\end{equation*}
\end{coroll}

\begin{propos}
  \label{prop:MeasurementMon}
  Both observational entropy and observational information quantity do not decrease when adding signals to the set $\Gamma$. Given a signal $\gamma \notin \Gamma$:
\begin{align*}
\IQd{\Gamma \cup\{\gamma\}}{d} & \ge \IQd{\Gamma}{d}\\
 \E{\Gamma\cup\{\gamma\}} & \ge \E{\Gamma}.
\end{align*}

\end{propos}

\begin{propos}
  \label{prop:FixLength}
  The observational entropy of a single ranking depends exclusively on its length. More formally, being $\gamma$ a  signal such that
  \begin{equation*}
  \gamma(d_1)>\gamma(d_2)>\ldots >\gamma(d_n) > \gamma(d_{n+1})=\gamma(d_{n+2})=\ldots,
  \end{equation*}
  then:
\begin{equation*} \E{\gamma}=-\frac{1}{|\D|}\sum_{i=1}^{n} log\left(\frac{i}{|\D|}\right).
\end{equation*}
\end{propos}

\begin{propos}
  \label{prop:redundancy}
  Observational entropy and observational information quantity  are  invariant under redundant signals. Being $f$ any strict  monotonic function (i.e., a function that does not affect the ordinal relationships)
\begin{align*}
\IQd{\Gamma\cup\{\gamma\}}{d} &= \IQd{\Gamma\cup\{\gamma, f(\gamma)\}}{d} \\
\E{\Gamma\cup\{\gamma\}} &= \E{\Gamma\cup\{\gamma, f(\gamma)\}}.
\end{align*}
\end{propos}

This aspect is crucial when representing documents in terms of systems output signals. Redundant systems should not increase the observational information quantity of documents.  In addition, adding a non-redundant signal  increases strictly the entropy of the signal set.

\begin{propos}
If a preference between two documents in $\gamma$ is not corroborated  by any signal in $\Gamma$, i.e.,
\begin{equation*}
\exists d_1, d_2 \in \D \ldotp \left( d_1 \geq_{\{\gamma\}} d_2 \wedge
d_1 \le_\Gamma d_2 \right),
\end{equation*}
then the entropy strictly increases when adding the signal to the set, i.e.,
\begin{equation*}
\E{\Gamma}< \E{\Gamma\cup\{\gamma\}}.
\end{equation*}
\end{propos}

\section{Measuring Effectiveness}
\label{sec:evaluation}

We now show how the above definitions and properties can be exploited to define an effectiveness measure that satisfies formal constraints which are not satisfied by traditional metrics.

\subsection{Observational Information Effectiveness}

As we said in the related work section, instead of  modeling the user behavior in the seeking process, we apply an information theory based similarity measure to compare system outputs against the gold-standard, but using our notions of observational information quantity and entropy. More specifically, we use the Information Contrast Model (ICM)~\cite{Amigo-ATIR-17}. ICM is a parameterizable extension of Point-wise Mutual Information which satisfies a set of constraints whenever its parameter are within a certain range~\cite{Amigo-ATIR-17}.
We start by defining a notion of effectiveness for a  signal.
\begin{definition}
  \textit{Given a  signal, $\gamma$, and a relevance gold standard, $g$, then the Observational Information Effectiveness (OIE) 
of the signal is a linear combination of observational entropies as follows:}
\begin{equation}
\label{eq:oie}
\eff(\gamma, g) = \alpha_1 \cdot \E{\gamma} + \alpha_2 \cdot \E{g}
                  - \beta \cdot \E{\{\gamma,g\}}.
\end{equation}
\end{definition}

where $\alpha_1, \alpha_2, \beta \in \mathbb{R}^+$ and $\alpha_1$ and $\alpha_2$ weight the effect of the ranking and assessment entropy. Hereafter, we will consider $\alpha_1=\alpha_2=1$ for the sake of simplicity. Note that the entropy of a single ranking with a fixed length is constant (Property \ref{prop:FixLength}). Therefore, when evaluating single rankings (i.e., when $\Gamma = \{r\}$) with a fixed length under a fixed gold, the parameters $\alpha_1$ and $\alpha_2$ do not affect the relative effectiveness of systems, which depends exclusively on the component $\E{\{\gamma,g\}}$.

\subsection{ Satisfying Formal Constraints}

We are interested in comparing the proposed metric with the state-of-the-art.
Comparing metrics empirically against user satisfaction or search effectiveness requires data that is often unavailable and expensive to collect. Moreover, findings may be biased to the subjects, retrieval systems or other experimental factors.

An alternative consists of studying evaluation metrics under formal constraints.
\citet{Amigo_2013} defined a theoretical framework according to
five formal constraints: 
swapping contiguous documents in concordance with the gold increases effectiveness (\emph{priority constraint}, \texttt{Pri}); 
the effect of swapping is larger at the top of the ranking (\emph{deepness constraint}, \texttt{Deep}); 
retrieving one relevant document is better than a huge amount of relevant documents after a huge set of irrelevant documents (\emph{deepness threshold constraint}, \texttt{DeepTh}); 
there exists a certain area at the top of the ranking in which $n$ relevant documents is better than only one (\emph{closeness threshold constraints}, \texttt{CloseTh});
and finally, adding irrelevant documents at the bottom of the ranking decreases effectiveness (\emph{confidence constraint}, \texttt{Conf}). 
According to this study, among the most popular metrics, only the  Rank-Biased Precision (RBP) metric~\cite{RBP} satisfies the  first four constraints. The following theorem states that OIE satisfies these five constraints.
\begin{theorem}
\label{Theo:MetricConstraint}
  {\bf Information Evaluation Theorem} OIE satisfies the five  constraints defined by \citet{Amigo_2013} whenever $1<\beta<\frac{2n-1}{n}$, being $n$  the minimum amount of documents that are necessarily explored by the user.\footnote{$n$ is a variable defined for the closeness deepness threshold constraint \cite{Amigo_2013}.}
\end{theorem}

Surprisingly, these theoretical boundaries for $\beta$ correspond with those predicted by~\citet{Amigo-ATIR-17} for the ICM similarity model, even though ICM is grounded on a different axiomatics, oriented to the general notion of similarity.

\subsection{Experiment}

Although this work is mainly theoretical, we performed a brief experiment 
comparing OIE against traditional metrics. Here, we use the meta-metric \emph{Metric Unanimity ($\mbox{MU}$)}~\cite{amigo2018axiomatic}. $\mbox{MU}$ quantifies to what extent a metric is sensitive to quality aspects captured by other existing metrics.  
The intuition is that, if a system improves another system for every quality criteria, this should be \emph{unanimously} reflected by every metric. A metric that captures every quality criteria should reflect these improvements.

$\mbox{MU}$ is formalized as the Point-wise Mutual Information (PMI) between metric decisions and improvements corroborated by all the metrics in a given set of metrics, $\mathcal{M}$. Formally, given a metric, $m \in \mathcal{M}$, and a set of system outputs, $\mathcal{R}$:\footnote{Note that the a priori probability of system improvement  for every metric is fixed $P(\Delta m_{i,j})=\frac{1}{2}$. That is, in the cases where two system outputs obtain the same score $m(s_i)=m(s_j)$, we  add $0.5$ to the statistical count.}
\begin{equation*}\label{eq:um}
\resizebox{.9\hsize}{!}{%
$\mbox{MU}_{\mathcal{M},\mathcal{R}}(m) = \mbox{PMI} \left( 
\Delta m_{i,j}, \Delta {\mathcal M_{i,j}} \right)
=
\log \left( 
   \frac
   {P(\Delta m_{i,j},\Delta {\mathcal M_{i,j}})}
   {P(\Delta m_{i,j})\cdot P(\Delta {\mathcal M_{i,j}})}     
    \right)$
    }.
\end{equation*}
In the equation, $\Delta m_{i,j}$ and $\Delta{\mathcal M}_{i,j}$  are statistical variables over 
system pairs $(r_i,r_j) \in \mathcal{R}^2$, indicating a system improvement  according to the metric and to every metric, respectively:
\begin{align*}
&\Delta m_{i,j}\equiv m(r_i)> m(r_j)\\
&\Delta{\mathcal M}_{i,j}\equiv\forall m\in \mathcal{M}\left(m(r_i)\ge m(r_j)\right).
\end{align*}

 \begin{table}[tbp]
 \caption{Traditional metrics and Observational Information  Effectiveness (OIE), ranked by Metric Unanimity (MU)~\cite{amigo2018axiomatic}. \tick\ indicates that the metric satisfies the formal constraint, \fail\ indicates otherwise.}
\begin{adjustbox}{width=0.48\textwidth,center}
  \centering
    \begin{tabular}{lcccccc}
    \toprule
    Metric & MU  & \texttt{Pri} & \texttt{Deep} & \texttt{DeepTh} & \texttt{CloseTh} & \texttt{Conf} \\
  \midrule
OIE$_{\beta=1.2}$ &{\bf $0.928$}  & \tick &\tick &\tick &\tick&\tick \\
\addlinespace
OIE$_{\beta=1}$ & $0.927$ & \tick &\tick &\tick &\fail&\fail \\
RBP &  $0.926$  & \tick &\tick &\tick &\tick&\fail \\
DCG  & $0.914$ &\tick &\tick &\fail &\tick&\fail \\
AP  & $0.910$ &\tick &\tick &\fail &\tick&\fail \\
P@$100$  & $0.910$ &\fail &\fail &\tick &\tick&\fail \\
DCG@$50$  & $0.905$ &\fail &\fail &\tick &\tick &\fail \\
ERR@$50$  & $0.903$ &\fail &\fail &\tick &\fail &\fail \\
ERR  & $0.901$ &\tick &\tick &\tick &\fail &\fail \\
P@$50$  & $0.900$ &\fail &\fail &\tick &\tick &\fail \\
ERR@$20$  & $0.886$ &\fail &\fail &\tick &\fail &\fail \\
DCG@$20$  & $0.886$ &\fail &\fail &\tick &\tick &\fail \\
P@$20$  & $0.876$ &\fail &\fail &\tick &\tick &\fail \\
P@$10$  & $0.829$ &\fail &\fail &\tick &\tick &\fail \\
RR@$10$  & $0.162$ &\fail &\fail &\tick &\fail &\fail \\
\bottomrule
  \end{tabular}
\label{tab:MU}
\end{adjustbox}
\end{table}

 Note that $\mbox{MU}$ is closely related with the unanimity notion in observational information quantity. The reader could thing that there exists some theoretical over-fitting here. However, $\mbox{MU}$ and observational information quantity ($\IQd{\Gamma}{d}$) measure different things. $\mbox{MU}_{\mathcal{M},\mathcal{R}}(m)$ measures the correspondence between a metric and the whole set of metrics (system rankings), while $\IQd{\Gamma}{d}$ combines system outputs (document rankings).
 
 In all our experiments, we  used the Gov-2 collection and the topics $701$ to $750$ used in the TREC 2004 Terabyte Track~\cite{clarke2004overview}. We also used the $60$ official runs submitted by the participants to the track. Table~\ref{tab:MU} shows the $\mbox{MU}$ results for OIE with $\beta=1.2$ (an arbitrarily selected value in the theoretical grounded range), and other standard evaluation metrics including: 
OIE with $\beta=1$ (out of the theoretically grounded  range), Precision at cutoff $k$ (P@$k$),
Average Precision (AP),
Reciprocal Rank (RR)~\cite{Voorhees99thetrec-8},
Expected Reciprocal Rank (ERR@$k$)~\cite{Chapelle-09},
Discounted Cumulative Gain (DCG@$k$)~\cite{DCG}
and
Rank-Biased Precision (RBP)~\cite{RBP}. For $\eff$, we have considered the cutoff point at position $100$. For the rest of metrics, we have considered the cutoff positions $20$, $50$ and $100$. For the RBP metric, we have considered the values $0.8$, $0.9$ and $0.99$ for the parameter $p$. 

As the table shows, metrics with shallow cutoffs ($20$ or $50$) and RR, which stops at the first relevant document, are at the bottom of the $\mbox{MU}$ ranking, given that they capture only partial aspects of the ranking quality.
On the other hand,  \eff$_{\beta=1.2}$  improves the rest of metrics in terms of $\mbox{MU}$. The improvement of \eff$_{\beta=1.2}$ regarding  \eff$_{\beta=1}$ corroborates the theoretical analysis about the $\beta$ ranges. 

Interestingly, RBP (the third ranked metric) is the only one that satisfies the four first constraints according to the study by \citet{Amigo_2013}.  Note that we obtained the same Unanimity for RBP regardless the $p$ parameter $(p\in\{0.8, 0.9, 0.99\})$. The improvement in $\mbox{MU}$ for \eff\ compared against RBP is probably due to the \textit{Confidence} constraint.
Some rankings have less than $100$ documents. The benefit of avoiding non-relevant documents at the end of the ranking is only rewarded by \eff\ in this metric set. 
Note also that some metrics such as DCG satisfy the Deepness Threshold constraint when adding a ranking cutoff, but this is at the cost of Priority and Deepness, given that documents at deep positions are not considered.

This experiment can  be extended for more metrics, meta-evaluation criteria, and data sets. In this paper we focus on the Observational Information Quantity as a theoretical framework that can be applied to different phenomena in IR. Our formal analysis shows that effectiveness can be grounded in the Observational Information framework. 

\section{Unsupervised Ranking Fusion}
\label{sec:fusion}

We now turn to applying the framework to the problem of ranking fusion.

\subsection{Information Quantity Cumulative Evidence}

 
 As we mentioned in the introduction, the experience in most of experiments reported in the IR literature corroborates the Probability Ranking Principle. For instance,  precision/recall curves tend to be descendant for every system. 
 Given that $\IQd{r}{d}$ is correlated with the  position of $d$ in the ranking $r$ (Property~\ref{prop:SingleSignal}), we can say that the probability of  relevance increases with the Observational Information Quantity under a single signal. 
We generalize this phenomenon for multiple signals stating the following assumption.
\begin{assum}
\label{IQEA}
[Information Quantity Cumulative Evidence] Adding signals increases the probability of improving relevance under an Observational Information Quantity increase
$$P\left(
       d\ge_g d_0 ~\big|~ 
          d\ge_{I_{\Gamma\cup\{\gamma\}}} d_0 
          \right)
\ge 
P\left(
    d\ge_g d_0 ~\big|~ d\ge_{I_{\Gamma}} d_0 
    \right).$$
\end{assum}

Note that according to Property~\ref{prop:SingleSignal}, the observational information quantity $\IQf{\gamma}$ under  a single signal $\gamma$ shorts documents in the same manner than $\gamma$. Therefore, we can directly infer from this theorem that an increase according to the $\IQf{\Gamma}$ is more reliable than an increase of signals in isolation. 

In order to check this assumption empirically, we make a pooling from the first $100$ ranked document for each system output in our data set. Here, we consider that documents not present in a particular ranking $\gamma$ are scored with the lowest signal value according to $\gamma$. 
In each experiment: 
\begin{enumerate}[(i)]
\item we randomly select one topic and a set $\Gamma$ of five system outputs; 
\item we compute the observational information quantity $\IQd{\Gamma}{d}$ for each document under this set of measurements $\Gamma$;
\item we select one single signal (one system output) $\gamma$ from $\Gamma$; 
\item 
We compute the conditional probabilities $P\left(d \ge_g d'| \gamma(d)\ge\gamma(d')\right)$ and $P\left(d \ge_g d'|d \ge_{\IQf{\Gamma}} d'\right)$. That is, the relevance increase,  when increasing the system score according to $\gamma$ and when increasing  the observational information quantity $\IQf{\Gamma}$. 
\end{enumerate}

\begin{figure}[tb]
  \begin{center}
      \centering
      \includegraphics[width=0.4\textwidth]{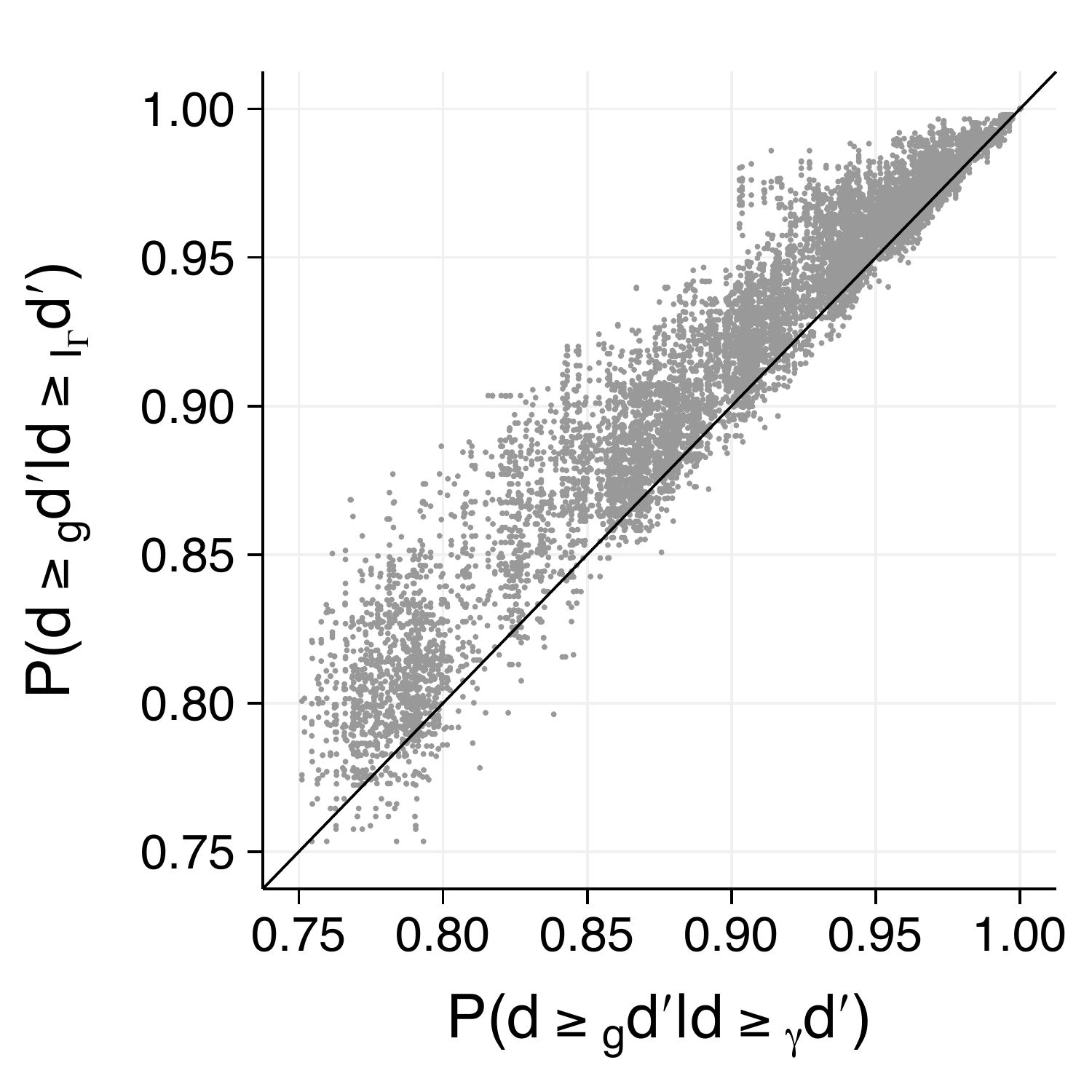}
      \caption{Checking the Information Effectiveness Additivity proposition.\label{fig:ExpIEA}}
  \end{center}
\end{figure}

Each dot in Figure \ref{fig:ExpIEA} represents one experiment, thus, one topic, five signals and one single signal from this set. The horizontal axis represents $P\left(d \ge_g d'|d \ge_\gamma d'\right)$. The vertical axis represents $P\big(d \ge_g d'|d \ge_{\IQf{\Gamma}} d'\big)$. As the figure shows, the Information Effectiveness Additivity proposition practically always holds.

\subsection{Ranking Fusion by Observational Information Quantity}

The ranking fusion model proposed in this paper consists of the combination of system outputs in a single signal according to the observational information quantity of documents. That is $\gamma_{\text{Fusion}}(d)=\IQf{\Gamma}(d)$.
Then, we can state the following theorem.

\begin{theorem}
\label{thRFOpt}
[Ranking Mergeability:] Under the Information Quantity Cumulative Evidence, and assuming that the information quantity estimation is fine grained, the effectiveness of $\IQf{\Gamma}$ for any $\beta$ values in the interval $(1,2)$ is higher than the effectiveness of any single measurement $\gamma\in\Gamma$:
$$ \eff_{\beta\in(1,2)} \left( \IQf{\Gamma \cup\{\gamma\}},g\right)
   \ge 
   \eff_{\beta\in(1,2)} \left(\IQf{\Gamma},g\right).$$
\end{theorem}

This theorem has strong practical implications. For instance, it means that instead of evaluating five systems, we can directly join them to achieve the best result. However, the Effectiveness Additivity Theorem has an important limitation, which is the need for high granularity in $\IQf{\Gamma}$. That is, we need a huge amount of documents and an extremely costly computation process to accurately estimate the probability of improvement for all measurements. In addition, there exist some theoretical limitations for the granularity in particular situations. For instance, two documents appearing at the first position of two different rankings have necessarily the same $\IQf{\Gamma}$, which is $\frac{1}{|\mathcal{D}|}$.

\begin{figure}[tb]
  \begin{center}
      \centering
      \includegraphics[width=0.4\textwidth]{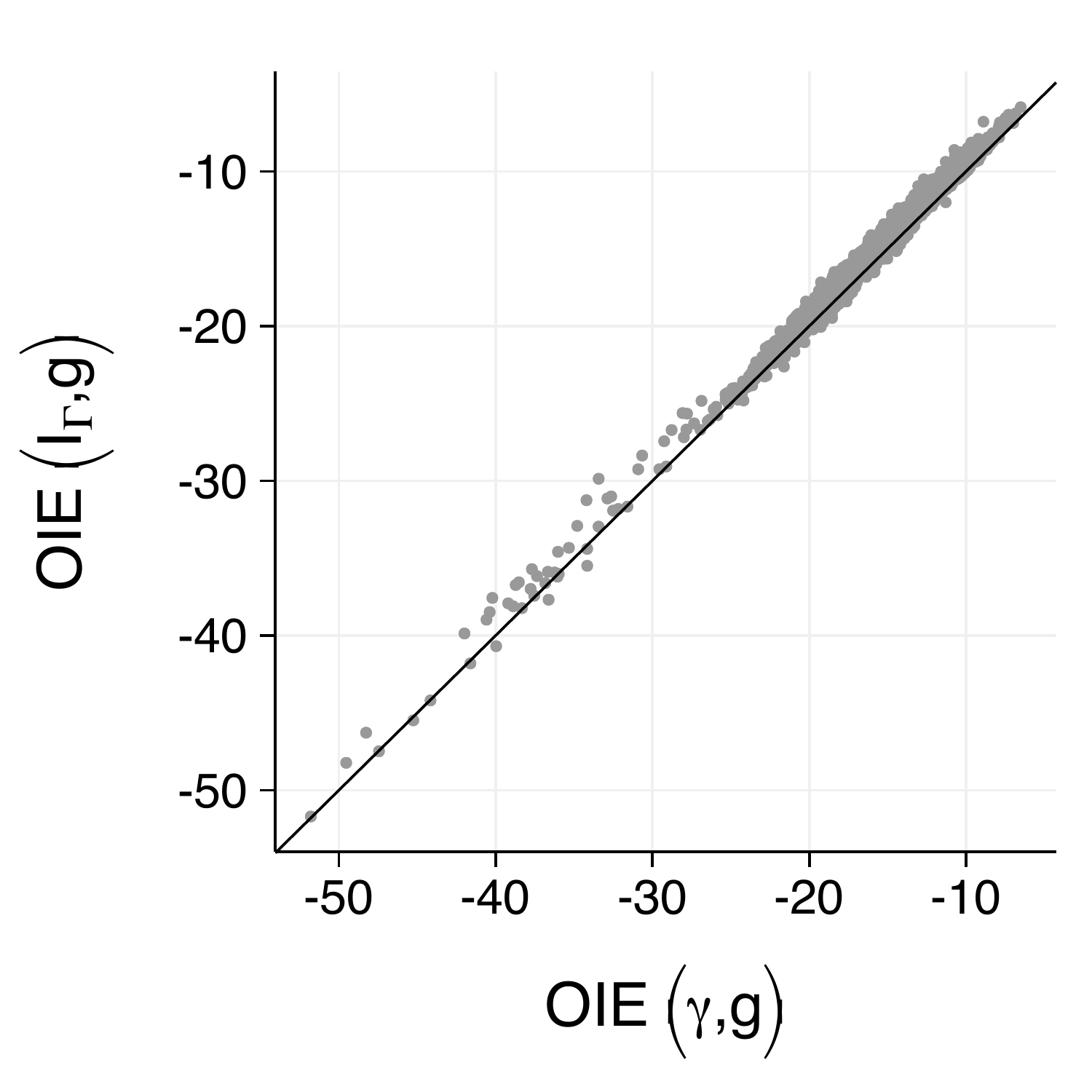}
      \caption{Experiment for checking the optimality of $\IQf{\Gamma}$ as ranking fusion method.\label{fig:ExpFusion}}
  \end{center}		
\end{figure}

Let us check the Ranking Mergeability theorem empirically. To this aim, we emulate fine-grained single signals and $\IQf{\Gamma}$ as follows. 
We start by generating random samples of five signals $\Gamma$, and one single signal $\gamma$ from $\Gamma$. 
Then, we collect documents from the single ranking $\gamma$ progressively from the top to the bottom, but discarding documents that achieve the same $\IQf{\Gamma}$ than previously collected documents.  This will generate a set of documents $\mathcal{D}'$ such that:
$$\forall d_1, d_2 \in \mathcal{D}'\ldotp  \IQd{\Gamma}{d_1}\neq 
\IQd{\Gamma}{d_2} \wedge 
\forall \gamma \in \Gamma \ldotp
\gamma(d_1)\neq \gamma(d_2).$$

Finally, we compare the effectiveness of $\gamma$ and $\IQf{\Gamma}$ 
in terms of OIE (using $\beta=1.2$)
across all topics, without considering the rest of documents. 

Figure \ref{fig:ExpFusion} compares the effectiveness of the single signal $\eff(\gamma,g)$ (x-axis) against the effectiveness of the combined signals $\eff(\IQf{\Gamma}, g)$ (y-axis). 
 In our experiments, $\eff(\IQf{\Gamma},g) > \eff(\gamma,g)$ for $1,809$ out of $2,000$ cases.
 
 As the theory predicts, $\IQf{\Gamma}$ outperforms the single signal in almost all the cases. Notice that the improvements are less prominent than in the previous experiment. The first reason is that we are considering rankings instead of probabilities under non strict comparisons between signal values. The second reason is that OIE captures \textit{top heaviness}, giving more weight to documents located at the top of rankings.

\subsection{Borda Count as an Approach to Observational Information Quantity}

An important drawback of the Observational Information Quantity as a ranking fusion method is the need for a huge amount of samples in the estimation process. The reason is that the probability of unanimous outscoring decreases dramatically when adding signals. A common way of estimating joint probabilities under a limited amount of data consists of assuming independence across variables $\left(P(A,B,C)\simeq P(A)\cdot P(B)\cdot P(C)\right)$. In this subsection we show that $\IQf{\Gamma}$ converges into a Borda Count variant consisting of averaging the logarithm of rank positions instead of the original ranking position. Note that Borda Count is the most popular way for averaging rankings. As described in Section~\ref{sec:theory}, our $\IQf{\Gamma}$ definition matches with the Information Quantity defined by \citet{Amigo-17} for similarity measure fusion. In their  paper it is proved that this definition  converges to other common ranking fusion algorithms depending on  statistical assumptions. In particular, they 
prove that:

\begin{proposition}
\label{prop:IQvsBorda}
Assuming statistical independence,  the Observational Information Quantity of a document under a set of rankings corresponds with the average logarithm of ranking positions: 
$$\IQd{\Gamma}{d}\propto \text{Avg}_{\gamma_i\in\Gamma} \log(\text{rank}_i(d)).$$
\end{proposition}

\begin{figure}[tb]
  \begin{center}
      \centering
            \includegraphics[width=0.4\textwidth]{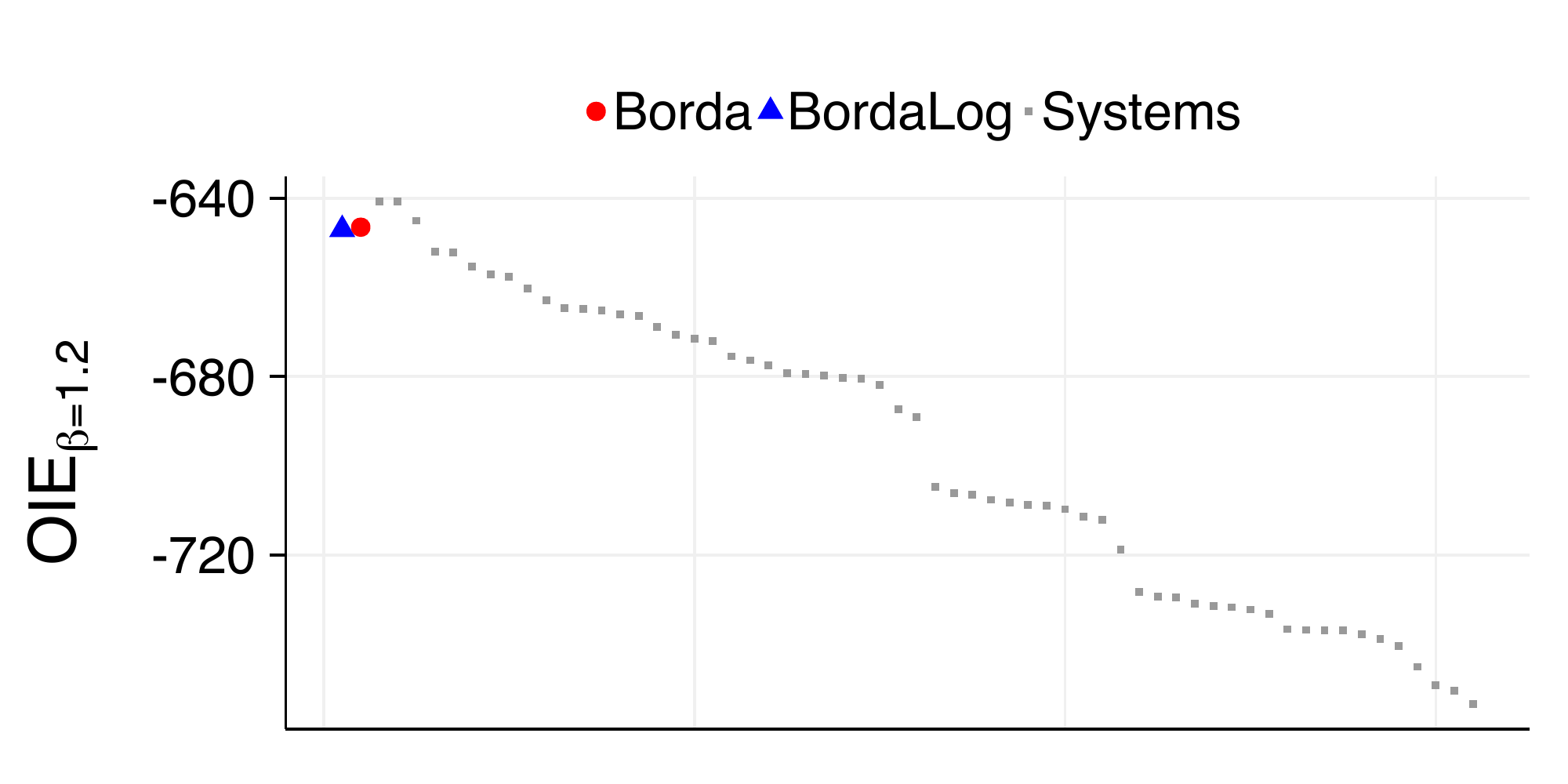}
      \caption{Evaluating the Borda fusion and the logarithmic Borda (BordaLog) over all system outputs. \label{fig:Borda_Eval}		}
  \end{center}
\end{figure}

Therefore, according to our analysis, in terms of Observational Information Effectiveness, if the independence assumption can be assumed, the Borda$_{\log}$ ranking fusion should be at least as effective than the best single system output. In addition, Borda$_{\log}$ should achieve a similar or better effectiveness than the original Borda algorithm.

Figure \ref{fig:Borda_Eval} illustrates the effectiveness of the original Borda, $\IQf{\Gamma}$ under the independence assumption (BordaLog) and each system in our experimental data set. We have used the $\beta$ value 1.2 according to the experiments in previous section. In order to evaluate outputs under the same conditions (fixed ranking length), we have truncated the Borda and BordaLog outputs in position $100$ just like single systems. The figure shows that, in concordance with the theoretical analysis, both fusion methods practically achieve the same performance as the best system in the combination.

\section{Connecting with Traditional  Information Theory}
\label{sec:derivation}
Probably, there are many possible theoretical explanations for the observational information framework. Signals (retrieval system outputs) in IR are quantitative, while the traditional information theory measures the information quantity of events characterized by binary features. On the other hand, \textit{Differential Entropy} considers a continuous space of signals, but it does not allow to estimate the information quantity of a single event in this continuous space. 

We now describe our proposed derivation for the observational information framework. We start by representing object observations as fuzzy sets. This allows us to capture both the amount of signals and their quantitative projection into each object. Then, we use the Dempster-Shafer theory of evidence~\cite{dempster1967upper,shafer1976mathematical}. In particular, we use the belief function over the fuzzy set operators in order to estimate the information quantity of observations. Note that the proposed model differs from other approaches based on Demster-Shafer theory which focus on document content representation~\cite{lalmas1998representing, teixeira1993belief}.

 A fuzzy set is formally defined as:

\begin{definition}
\label{def:FuzzySet}
\textit{A fuzzy set is a pair $(A,f)$ where $A$ is a set and $f$ a membership function $f:A\longrightarrow[0,1]$.}
\end{definition}

Then, an  observation can be formalized as follows.

\begin{definition}
\label{def:Observation}
Given a set of signals, $\Gamma=\{\gamma_1,\ldots,\gamma_n\}$, and a set of possible values generated by each signal, $\{ x_1, \ldots, x_n \}$, an observation, $\obsd{\Gamma}{x_1,\ldots,x_n}$, under $\Gamma$ is a fuzzy set of signals whose  membership function corresponds to the signal values: $A=\Gamma$ and $f(\gamma_i)=x_i$, $\forall i \in \{1,\ldots, n\}$.
\end{definition}

In other words, a document observation has two main components: the signals under which the document is observed and the corresponding signal values. From the previous definition we can infer that each document, $d \in \D$, produces an observation,  $\obsd{\Gamma}{\gamma_1(d),\ldots,\gamma_n(d)}$, denoted as $\obsd{\Gamma}{d}$. 


According to the inclusion operator in fuzzy sets, an observation is included into another when:
$$\obsd{\Gamma}{x_1, \ldots, x_n}\subseteq\obsd{\Gamma}{x'_1, \ldots, x'_n}\Longleftrightarrow \forall i=1,\ldots, n; x'_i\ge x_i.$$

The purpose of Dempster-Shafer's theory is to represent believes in a set of elements referred to as a \emph{frame of discernment}. We can consider the \emph{believe function} defined on the observations of documents, this is possible by taking into account the inclusion relationship between observations.

Then, applying the Dempster-Shafer evidence theory, we define the mass function  or \emph{Basic Probability Assignment (BPA)} as the probability of observations across the set of documents $\D$. Being  $\omega$ an observation:
$$m(\omega)=P_{d \in\D}\big(\obsd{\Gamma}{d}=\omega \big).$$


Consequently, the corresponding \textit{belief function} of an observation $\omega'$ is:
$$Bel(\omega')=\sum_{\omega|\omega\supseteq \omega'} m(\omega).$$
And therefore, the belief of a document observation can be expressed as:
$$
Bel(\obsd{\Gamma}{d})=\sum_{\omega|\omega\supseteq \obsd{\Gamma}{d}} P_{d' \in \D}\big(\omega=\obsd{\Gamma}{d'} \big)=P_{d' \in \D}\big(\obsd{\Gamma}{d'}\supseteq \obsd{\Gamma}{d}\big).$$


The last identity is attained by considering that observations are actually a partition of the document space.

Then, the observational information quantity is analogous to the information quantity in Shannon's theory but replacing the probability with the belief function:
\begin{align*}
I_{\Gamma}(d)&=I(\obsd{\Gamma}{d})=-\log(Bel(\obsd{\Gamma}{d}))\\
&=-\log(P_{d'\in\D}(\obsd{\Gamma}{d'}\supseteq \obsd{\Gamma}{d}))\\
&=-\log(P_{d'\in\D}(d'\ge_\Gamma d)),
\end{align*}
which leads directly to Definition~\ref{def:FPIQ}.


The entropy of a set of signals is directly derived from the traditional notion. That is, the expected information quantity:
\begin{align*}
H(\Gamma)&=\sum_{x_1}\cdots \sum_{x_n} P(x_1,\ldots,x_n) I(\obsd{\Gamma}{(x_1,\ldots,x_n)}\\
&= \sum_{\omega\in\Omega_{\D}} P_{d\in\D }(\omega=\obsd{\Gamma}{d}) I_\Gamma(d)=\frac{1}{\D}\sum_{d\in\D} I_\Gamma(d),
\end{align*}
which leads to Definition~\ref{def:H}.

\section{Conclusions}
\label{sec:conclusions}
We have shown how --by starting from the  Shannon-like definitions of Observational Information Quantity and Observational Entropy-- we can provide a theoretically grounded explanation of phenomena that are well known results of empirical experiments. In this paper we have focused on effectiveness metrics and ranking fusion. Effectiveness can be modeled in terms of information theory --Observational Information Effectiveness (OIE), which is based on the similarity between system outputs and human assessments. OIE satisfies desirable properties that are not satisfied by traditional metrics. Moreover, our experimental results suggest that OIE captures aspects from different existing metrics. Regarding the ranking fusion problem, we have seen that, under certain assumptions, the observational information quantity outperforms single signals. 

This current work has known limitations, given that the estimation of observational information quantity is not straightforward. In the near-future we plan to apply our general framework to explain other phenomena that are important in Information Retrieval, such as evaluation without relevance assessment and query performance prediction.

\section*{Acknowledgements}
This research was partially supported 
by the
Spanish Government (project Vemodalen TIN2015-71785-R)
and the Australian Research Council (project LP150100242).

\bibliographystyle{ACM-Reference-Format}
\bibliography{strings-shrt,bibliography}

\appendix
\newpage
\section{Formal Proofs}
\label{sec:appendix}

\begin{proof}
{\small
{\bf [Property \ref{prop:ValMon}]}\\
According to 
Definition~\ref{def:FPIQ}, if 
$\forall \gamma \in \Gamma \left(\gamma(d_1)\ge \gamma(d_2)\right)$ then:
$$\IQd{\Gamma}{d_1}=-\log\left(P_{d'\in \mathcal{D}}\big( d' \ge_{\Gamma} d_1\big)\right) \ge$$
$$-\log\left(P_{d' \in \mathcal{D}}\big( d' \ge_{\Gamma} d_2\big)\right) =
\IQd{\Gamma}{d_2} $$
}
\end{proof}

\begin{proof}
{\small 
{\bf [Property \ref{prop:MeasurementMon}]}\\
According to 
Definition~\ref{def:FPIQ}:
$$\IQd{\Gamma \cup\{\gamma\}}{d}=-\log\left(P_{d' \in \mathcal{D}}\big( d'\ge_{\Gamma \cup \{\gamma\}} d\big)\right) 
\ge $$
$$-\log\left(P_{d' \in \mathcal{D}}\big( d'\ge_{\Gamma} d\big)\right)=\IQd{\Gamma}{d}$$
And therefore:
$$H(\Gamma\cup\{\gamma\})=\frac{\sum_{d\in \D } \IQd{\Gamma\cup \{\gamma\}}{d}}{|\D|}\ge\frac{\sum_{d\in \D} \IQd{\Gamma}{d}}{|\D|}=
H\big(\Gamma\big)$$
}
\end{proof}

\begin{proof}
{\small 
{\bf [Property \ref{prop:FixLength}]}\\
$$ H(\{\gamma\})=\frac{\sum_{d\in\D} \IQd{\{\gamma\}}{d}}{|\D|}=\frac{\sum_{d\in\D} -\log\left(P_{d'\in \mathcal{D}}\big( d'\ge_{\{\gamma\}} d\big)\right)}{|\D|}=$$
$$\frac{1}{|\mathcal{D}|}\sum_{i=1}^{n} \log\left(\frac{|\mathcal{D}|}{i}\right)$$
}
\end{proof}

\begin{proof}
{\small 
{\bf [Property \ref{prop:redundancy}]}\\
Being $f$ any strict  monotonic function (i.e. does not affect the ordinal relationships)

$$\IQd{\Gamma\cup\{\gamma\}}{d} = 
-\log\left(P_{d'\in \mathcal{D}}\big( d'\ge_{\Gamma \cup \{\gamma\}} d\big)\right) =$$
$$-\log\left(P_{d' \in \mathcal{D}}\big( d'\ge_{\Gamma \cup\{\gamma,f(\gamma)\}} d\big)\right) =
\IQd{\Gamma\cup\{\gamma,f(\gamma)\}}{d}$$
and therefore:
$$ H\big(\Gamma\cup\{\gamma\}\big)= H\big(\Gamma\cup\{\gamma, f(\gamma)\}\big)$$
}
\end{proof}

\begin{proof}
{\bf [Theorem \ref{Theo:MetricConstraint} (Observational Information Evaluation Theorem)]:}
{\small
Let be $\mathcal{D}$ the collection of documents. For our purposes, we can ignore the $H(g)$ from $\eff$ given that it affects equally to every compared outputs.

Regarding the priority and deepness constraints, when swapping two contiguous documents in the ranking in concordance with the gold ($g(d_i)=0,g(d_{i+1})=1$):
\begin{align*}
&\eff(r_{d_i\leftrightarrow d_{i+1}})-\eff(r)=H(r_{d_i\leftrightarrow d_{i+1}})-\beta H(\{r_{d_i\leftrightarrow d_{i+1}},g\})-H(r)+\beta H(\{r,g\})\\
&=-\beta H(\{r_{d_i\leftrightarrow d_{i+1}},g\})+\beta H(\{r,g\})\propto 
- H(\{r_{d_i\leftrightarrow d_{i+1}},g\})+ H(\{r,g\})\\
&\propto\sum_{d\in\mathcal{D}} \IQd{\{r,g\}}{d}-\sum_{d\in\mathcal{D}} \IQd{\{r_{d_i\leftrightarrow d_{i+1}},g\}}{d}\\
&=\IQd{\{r,g\}}{d_i}+\IQd{\{r,g\}}{d_{i+1}}-
\IQd{\{r_{d_i\leftrightarrow d_{i+1}},g\}}{d_i}-\IQd{\{r_{d_i\leftrightarrow d_{i+1}},g\}}{d_{i+1}} 
\end{align*}
 Given that in both rankings the amount of relevant documents above the relevant document $d_{i+1}$ is equal, and therefore:
 $$\IQd{\{r,g\}}{d_{i+1}}=
 \IQd{\{r_{d_i\leftrightarrow d_{i+1}},g\}}{d_{i+1}}$$
Therefore, the previous expression is equivalent to:
\begin{align*}
&\IQd{\{r,g\}}{d_i}-\IQd{\{r_{d_i\leftrightarrow d_{i+1}},g\}}{d_i}\\
&=-\log\left(P_{d}(d\ge_{\{g,r\}} d_i \right))
 +\log(P_{d}(d\ge_{\{g, r_{d_i\leftrightarrow d_{i+1}}\}} d_i))\\
 &=-\log\left(\frac{i}{|\mathcal{D}|}\right)+\log\left(\frac{i+1}{|\mathcal{D}|}\right)=\log\left(\frac{i+1}{i}\right)
\end{align*}
Given that $\log(\frac{i+1}{i})$ is positive and monotonic regarding $i$, both the priority and deepness constraints are satisfied. 

Regarding the threshold constraints, being $D_g$ the set of documents annotated as relevant in the gold standard. Let $r_1$ a ranking which retrieves only one relevant document, its $\eff$ is:
\begin{align*}
\eff(r_1,g)&\simeq H(r_1)-\beta H(\{r_1,g\})\\
&\propto \IQd{\{r_1\}}{d}-\beta\left(\IQd{\{r_1,g\}}{d}+\sum_{d'\in D_g-\{d\}}\IQd{\{r_1,g\}}{d'}\right)\\
 &=-\log\left(\frac{1}{|\mathcal{D}|}\right)-\beta\left(-\log\left(\frac{1}{|\mathcal{D}|}\right)-
 \sum_{d'\in D_g -\{d\}}\left(\log\left(\frac{|D_g|}{|\mathcal{D}|}\right)\right)\right)=\\
 &=(1-\beta)\left(-\log\left(\frac{1}{|\mathcal{D}|}\right)\right)-
 	\beta(|D_g|-1)\left(-\log\left(\frac{|D_g|}{|\mathcal{D}|}\right)\right)
\end{align*}

On the other hand, let $r_n$ a ranking which retrieves $n$ relevant documents after $n$ non relevant documents, then, $H(r_n)$ is $-\sum_{i=1}^{2n}\log\left(\frac{i}{|\mathcal{D}|}\right)$, and $H(\{r_n,g\})$ is proportional to:
\begin{align*}
&\sum_{i=1}^n-\log\left(\frac{i}{|\mathcal{D}|}\right)
+\sum_{i=n+1}^{2n}-\log\left(\frac{i-n}{|\mathcal{D}|}\right)
+(|D_g|-n)\left(-\log\left(\frac{|D_g|}{|\mathcal{D}|}\right)\right)
\\
&=-2\sum_{i=1}^{n}\log\left(\frac{i}{|\mathcal{D}|}\right)
+(N_g-n)\left(-\log\left(\frac{|D_g|}{|\mathcal{D}|}\right)\right)
\\
\end{align*}
Therefore, $\eff(r_n, g)=H(r_n)-\beta H(\{r_n,g\})$ can be expressed as:
\begin{align*}
-&\sum_{i=1}^{2n}\log\left(\frac{i}{|\mathcal{D}|}\right)-\beta\left(-2\sum_{i=1}^{n}\log\left(\frac{i}{|\mathcal{D}|}\right)
+(N_g-n)\left(-\log\left(\frac{|D_g|}{|\mathcal{D}|}\right)\right)\right)\\
=&-\sum_{i=n+1}^{2n}\log\left(\frac{i}{|\mathcal{D}|}\right)+(2\beta-1) \sum_{i=1}^{n}\log\left(\frac{i}{|\mathcal{D}|}\right)-\beta(|D_g|-n)\left(-\log\left(\frac{|D_g|}{|\mathcal{D}|}\right)\right)\\
\end{align*}
In order to satisfy the \textit{Deepness Threshold} constraint, the effectiveness $\eff(r_n, g)$ should tend to
$-\infty$ then n is extremely large. Then:
\begin{align*}
\lim_{n\rightarrow \infty}-\sum_{i=n+1}^{2n}&\log\left(\frac{i}{|\mathcal{D}|}\right)+(2\beta-1) \sum_{i=1}^{n}\log\left(\frac{i}{|\mathcal{D}|}\right)\\
&-\beta(|D_g|-n)\left(-\log\left(\frac{|D_g|}{|\mathcal{D}|}\right)\right)\le\\
\lim_{n\rightarrow \infty}-\sum_{i=n+1}^{2n}&\log\left(\frac{i}{|\mathcal{D}|}\right)+(2\beta-1) \sum_{i=1}^{n}\log\left(\frac{i}{|\mathcal{D}|}\right)<\\
\lim_{n\rightarrow \infty}-\sum_{i=1}^{n}\log&\left(\frac{i}{|\mathcal{D}|}\right)+(2\beta-1) \sum_{i=1}^{n}\log\left(\frac{i}{|\mathcal{D}|}\right)=\\
\lim_{n\rightarrow \infty}(2\beta-2)& \sum_{i=1}^{n}\log\left(\frac{i}{|\mathcal{D}|}\right)=
-\infty \mbox{~~whenever~~}\beta>1\\
\end{align*}

In order to satisfy \textit{Closeness Deepness} constraint:
\begin{align*}
&\eff(r_1,g)<\eff(r_n,g)\leftrightarrow
(1-\beta)\left(-\log\left(\frac{1}{|\mathcal{D}|}\right)\right)-\beta(N_g-1)\left(-
\log\left(\frac{|D_g|}{|\mathcal{D}|}\right)\right)<\\
&-\sum_{i=n+1}^{2n}\log\left(\frac{i}{|\mathcal{D}|}\right)+(2\beta-1) \sum_{i=1}^{n}\log\left(\frac{i}{|\mathcal{D}|}\right)-\beta(|D_g|-n)\left(-\log\left(\frac{|D_g|}{|\mathcal{D}|}\right)\right)
\end{align*}

Assuming that $\beta>1$ and $N_g>n$, we  need to prove that:
\begin{align*}
\eff(r_1,g)<\eff(r_n, g)&\leftarrow\\
(1-\beta)&\left(-\log\left(\frac{N_g}{|\mathcal{D}|}\right)\right)-\beta(N_g-1)\left(-
\log\left(\frac{|D_g|}{|\mathcal{D}|}\right)\right)<\\
&-\sum_{i=n+1}^{2n}\log\left(\frac{i}{|\mathcal{D}|}\right)+(2\beta-1) \sum_{i=n+1}^{2n}\log\left(\frac{i}{|\mathcal{D}|}\right)\\
&-\beta(|D_g|-n)\left(-\log\left(\frac{|D_g|}{|\mathcal{D}|}\right)\right)\Leftrightarrow\\
(1-\beta)&\left(-\log\left(\frac{|D_g|}{|\mathcal{D}|}\right)\right)-\beta(N_g-1)\left(-
\log\left(\frac{|D_g|}{|\mathcal{D}|}\right)\right)<\\
&-\sum_{i=n+1}^{2n}\log\left(\frac{|D_g|}{|\mathcal{D}|}\right)+(2\beta-1) \sum_{i=n+1}^{2n}\log\left(\frac{|D_g|}{|\mathcal{D}|}\right)\\
&-\beta(|D_g|-n)\left(-\log\left(\frac{|D_g|}{|\mathcal{D}|}\right)\right)\Leftrightarrow\\
(1-\beta)&-\beta(|D_g|-1)<
n+(1-2\beta)n-\beta(|D_g|-n)\Leftrightarrow\\
&\beta(-1-|D_g|+1+2n+|D_g|-n)<2n-1\Leftrightarrow\beta<\frac{2n-1}{n}
\end{align*}

Finally, regarding the \textit{Confidence} constraint, 
when adding a non-relevant document $d$ in the last position of $r$, the effect is that $d$ is the only document which Observational Information Quantity changes.
In addition, a non-relevant document according to $g$ satisfies 
$$\forall d' \in \mathcal{D}\ldotp g(d') \ge g(d) $$
which implies that $\IQd{\{r,g\}}{d}=\IQd{\{r\}}{d}$. 
Therefore, we can say that the increase of $H(\{r,g\})$ is equal than the increase of $H(\{r\})$. Therefore, according to the $\eff$ definition, the score decreases whenever $\beta > \alpha_1$, satisfying \textit{Confidence}.

}
\end{proof}



\begin{proof}
{\bf [Theorem \ref{thRFOpt}:Ranking Mergeability] }
{\small
\begin{align*}
&\eff\big(\IQf{\Gamma\cup\{\gamma\}}\big)\ge \eff\big(\IQf{\Gamma}\big)\Longleftrightarrow\\
&H(\IQf{\Gamma\cup\{\gamma\}})- \beta H(\IQf{\Gamma\cup\{\gamma\}}\cup\{g\}) \ge 
 H(\IQf{\Gamma})- \beta H(\IQf{\Gamma}\cup\{g\})
\end{align*}

Assuming that $\IQf{\Gamma\cup\{\gamma\}}$ and $\IQf{\Gamma}$ are fine grained then:
$$H(\IQf{\Gamma\cup\{\gamma\}})=H(\IQf{\Gamma})$$
Therefore:
\begin{align*}
&H(\IQf{\Gamma\cup\{\gamma\}})- \beta H(\IQf{\Gamma\cup\{\gamma\}}\cup\{g\}) \ge 
 H(\IQf{\Gamma})- \beta H(\IQf{\Gamma}\cup\{g\})\Longleftrightarrow\\
&- \beta H(\IQf{\Gamma\cup\{\gamma\}}\cup\{g\}) \ge - \beta H(\IQf{\Gamma}\cup\{g\})\Longleftrightarrow\\
&-  H(\IQf{\Gamma\cup\{\gamma\}}\cup\{g\}) \ge - H(\IQf{\Gamma}\cup\{g\})\Longleftrightarrow\\
&\prod_{d_0} P_d\big(d\ge_{I_{\Gamma\cup\{\gamma\}}}d_0,d\ge_g d_0\big)\ge
\prod_{d_0} P_d\big(d\ge_{I_{\Gamma}}d_0,d\ge_g d_0\big)\Longleftrightarrow\\
&\prod_{i=1..n}\frac{i}{n}\prod_{d_0} P_d\big(d\ge_{I_{\Gamma\cup\{\gamma\}}}d_0,d\ge_g d_0\big)\ge
\prod_{i=1..n}\frac{i}{n}\prod_{d_0} P_d\big(d\ge_{I_{\Gamma}}d_0,d\ge_g d_0\big)\Longleftrightarrow\\
&\prod_{d_0} \frac{P_d\big(d\ge_{I_{\Gamma\cup\{\gamma\}}}d_0,d\ge_g d_0\big)}
{P_d\big(d\ge_{I_{\Gamma\cup\{\gamma\}}}d_0\big)} \ge
\prod_{d_0} \frac{P_d\big(d\ge_{I_{\Gamma}}d_0,d\ge_g d_0\big)}
{P_d\big(d\ge_{I_{\Gamma}}d_0\big)}\Longleftrightarrow\\
&\prod_{d_0} P_d\big(d\ge_g d_0~\big|~d\ge_{I_{\Gamma\cup\{\gamma\}}}d_0\big)\ge
\prod_{d_0} P_d\big(d\ge_g d_0~\big|~d\ge_{I_{\Gamma}}d_0\big)
\end{align*}
which is true according to the Information Quantity Cumulative Evidence assumption.
}
\end{proof}

\end{document}